\documentstyle[epsfig]{article}

\newbox\grsign \setbox\grsign=\hbox{$>$}  
\newdimen\grdimen \grdimen=\ht\grsign  
\newbox\laxbox \newbox\gaxbox  
\setbox\gaxbox=\hbox{\raise.5ex\hbox{$>$}\llap  
     {\lower.5ex\hbox{$\sim$}}}\ht1=\grdimen\dp1=0pt  
\setbox\laxbox=\hbox{\raise.5ex\hbox{$<$}\llap  
     {\lower.5ex\hbox{$\sim$}}}\ht2=\grdimen\dp2=0pt  
\def\gax{\mathrel{\copy\gaxbox}}  
\def\lax{\mathrel{\copy\laxbox}}  
\def\lta{\lax}  
\def\gta{\gax}  
  
\begin{document}

\title{ADAFs -- Models, Observations and Problems}
\author{Jean-Pierre Lasota\\
{\it DARC, Observatoire de Paris, 92190 Meudon, France}\\
{\it lasota@obspm.fr}\\}
\date{}
\maketitle

\begin{abstract}
We review some of the properties of Advection-Dominated Accretion Flow
(ADAF) models and show that they successfully describe many
astrophysical systems. Despite these successful applications some
fundamental problems still remain to be solved, the most important one
being the physics of the transition between an ADAF and a geometrically
thin Keplerian disc.
\end{abstract}

\noindent
\leftline{PACS: 04.70.Bw; 97.60.Lf}
\leftline{Keywords: Black holes; Accretion discs}

\section{Introduction}

`Advection Dominated Accretion Flows' (ADAFs) is a term describing infall
of matter with angular momentum, in which radiation efficiency is very
low. In their applications, ADAFs are supposed to describe inflows onto
compact bodies, such as black holes or neutron stars; but very hot,
optically thin flows are  bad radiators in general so that, in
principle, ADAFs are possible in other contexts. Of course in the
vicinity of black holes or neutron stars, the virial (gravitational)
temperature is $T_{\rm vir}\approx 5 \times 10^{12} r^{-1}$ K, where
$r$ is the radial distance measured in the units of the Schwarzschild
radius $R_{\rm S}= 2GM/c^2$, so that in optically thin plasmas, at such
temperatures, both the coupling between ions and electrons and the
efficiency of radiation processes are rather feeble. In
such a situation, the thermal energy released in the flow by the
viscosity, which drives accretion by removing angular momentum, is not
going to be radiated away, but will be {\sl advected} towards the
compact body. If this compact body is a black hole, the heat will be
lost forever, so that advection, in this case, acts like a `global'
cooling mechanism.  In the case of infall onto a neutron star, the
accreting matter lands on the star's surface and the (reprocessed)
advected energy will be radiated away. There, advection may act only as
a `local' cooling mechanism. (One should keep in mind that, in
general, advection may also be responsible for heating, depending on
the sign of the temperature gradient - for example, according to
Nakamura et al.  (1997), in some conditions, near the black hole,
advection heats up electrons in a two-temperature ADAF).

The role of advection in an accretion flow depends on the radiation
efficiency, so that it depends on the microscopic state of  matter and
on the absence or presence of a magnetic field. If, for a given
accretion rate, radiative cooling is not efficient, advection is
necessarily dominant, assuming that a stationary solution is possible
(see below).

\subsection{Simple ADAF solutions}

One can illustrate fundamental properties of ADAFs by a simple example.
The advection `cooling'  (per unit surface) term in the energy equation 
can be written as
\begin{equation}
Q_{\rm adv} = {\dot M \over 2 \pi R^2} c_{\rm s}^2 \xi
\label{advterm}
\end{equation}
where $\dot M$ is the accretion rate, $c_{\rm s}$ the (adiabatic)
speed of sound and 
$\xi = -\left[(\gamma-1)^{-1} (d \ln T/d \ln R) - (d
\ln \Sigma/ d {\ln R})\right] $ is the dimensionless advection factor  
characterizing the entropy gradient (Abramowicz et al. 1995). $\gamma$ is
the ratio of specific heats and we assume that the pressure is provided only
by gas.

Using the hydrostatic equilibrium equation
\begin{equation}
{H \over R } \approx {c_{\rm s} \over v_{\rm K}}
\label{hydroeq}
\end{equation}
where $H$ is the disc semi-thickness and $v_{\rm K}= \sqrt{GM/R} 
\equiv c/\sqrt{2r}$ 
the  Keplerian velocity, one can write the advection term as
\begin{equation}
Q_{\rm adv} = {\dot m \xi\over 2 \kappa_{\rm es}R_{\rm S}}
               \left({c \over r}\right)^3  
               \left({H \over R }\right)^2 
\label{advterm2}
\end{equation}
whereas the viscous heating term can be written as
\begin{equation}
Q_{\rm vis} = {3 \dot m\over 8 \kappa_{\rm es}R_{\rm S}}
               \left({c \over r}\right)^3
\label{visheat} 
\end{equation}
where $\dot m = (\dot M c \kappa_{\rm es}/ 4\pi GM)$ and $\kappa_{\rm es}$
is the electron-scattering opacity coefficient. Therefore,
for very high temperatures, when $(H/R) \sim 1$ the advection
term is comparable to the viscous term and cannot be neglected.

From Eqs. (\ref{advterm2}) and (\ref{visheat}) one can easily obtain
an advection dominated solution by writing 
\begin{equation}
 Q_{\rm vis}= Q_{\rm adv}
\end{equation} 
and using 
\begin{equation} 
\left({H \over R }\right)^2 =
              {\sqrt 2 \over \kappa_{\rm es}} \dot m 
                     \left(\alpha \Sigma\right)^{-1}r^{-1/2}. 
\label{bizarre} 
\end{equation}
An ADAF is then described by the relation 
\begin{equation} 
\dot m = 0.53 \kappa_{\rm es} \alpha r^{1/2} \Sigma.  
\end{equation} 
In Eq. (\ref{bizarre}) we use the mass and angular momentum
conservation equations and the viscosity prescription $\nu=(2/3)c_{\rm
s}^2/\Omega_K$ ($\Omega_K =v_{\rm K}/R$).

\subsection{Maximum accretion rate}

Of course, the energy equation is
\begin{equation}
 Q_{\rm vis}= Q_{\rm adv} + Q_{\rm rad}
\label{energy}
\end{equation}
where $Q_{\rm rad}$ is the radiative cooling (per unit surface), so that
the existence of ADAF solutions depends on $Q_{\rm rad}$. There
is no universal form for this term, which depends on the microscopic
state of the matter, its content and the optical thickness.
In the simplest case of an optically thin, one-temperature flow, 
cooled by non-relativistic free-free processes,
\begin{equation}
Q_{\rm rad} \sim \rho^2 T^{3/2} \sim \alpha^{-2} \dot m^2 r^{-2}
\label{rad}
\end{equation}
so that the cooling term has a steeper dependence on $\dot m$ than
both $ Q_{\rm vis}$ and $Q_{\rm adv}$.  One can see therefore that there
exists a maximum value of $\dot m$ (Abramowicz et al. 1995) for which
an ADAF solution is possible:  $\dot m \sim \alpha^{2} r^{-1/2}$. We have 
assumed here that  Eq. (\ref{rad}) applies for this value of $\dot m$
(see below).

Abramowicz et al. (1995) found, in the non-relativistic free-free cooling case, 
that the maximum accretion rate for an ADAF solution is
\begin{equation}
\dot m_{\rm max} = 1.7 \times 10^3 \alpha^2 r^{-1/2}.
\label{mdotmax}
\end{equation}
The value of $\dot m_{\rm max}$ depends on the cooling in the flow and 
non-relativistic free-free cooling is not a realistic description of
the emission in the vicinity ($r\lta 10^3$) of a black hole. The flow there
will most probably form a two-temperature plasma (Narayan \& Yi 1995).  
More realistic calculations in a 2T flow by
Esin et al. (1997) give $\dot m_{\rm max}\approx 10 \alpha^2$ with
almost no dependence on radius. For larger radii $\dot m_{\rm max}$
decreases with radius.

\subsection{Maximum $\alpha$}

The existence of a maximum accretion rate for an ADAF is, as explained above,
due to the form of the cooling law. For radii $\lta 10^3$ optically thin cooling
of gas pressure dominated plasma implies the existence of such a critical
accretion rate. If, however, effects of optical thickness and of radiative
pressure become important, the slope of $Q_{\rm rad}(\dot m)$ will change
and, as shown by Chen et al. (1995), one may get families of solutions in
which an ADAF is possible for all values of $\dot m$
because $Q_{\rm rad}(\dot m)$ has a maximum
for $Q_{\rm rad} < Q_{\rm adv}$. The existence of such solutions depends on
the value of $\alpha$: $\dot m_{\rm max}$ exists only for 
$\alpha < \alpha_{\rm max}$, where the value of $\alpha_{\rm max}$ depends
on the cooling mechanism and radius. As a simple example one can use a
1T disc and consider that the simple free-free cooling formula is not valid 
for $\tau_{\rm eff}\gta 1$, where the effective optical depth 
$\tau_{\rm eff}=
\sqrt{\kappa_{\rm ff}(\kappa_{\rm ff}+\kappa_{\rm es})} \Sigma /2$ and 
$\kappa_{\rm ff}$ is the free-free opacity. Using a Planck mean value for
this optical depth, one obtains from $\dot m_{\rm max}< 
\dot m (\tau_{\rm ff}=1)$
a condition for the viscosity parameter:
\begin{equation}
\alpha < \alpha_{\rm crit} \approx r
\label{acrit}
\end{equation}
Chen et al. (1995) obtained values of $\alpha_{\rm crit}$ between 0.2 
and 0.4 for $r=5$, depending on the model, and Bj\"ornsson et al. (1996)
get  $\alpha_{\rm crit}> 1$. If the viscous stress is assumed to 
be proportional to the gas, and not to the total, pressure,
the value of $\alpha_{\rm crit}>>1$ (L\o v\aa s 1998) and is of no
physical interest.
It seems that for physically interesting
configurations ($\alpha \lta 1$) there is always a $\dot m_{\rm max}$
for ADAFs but one should not forget that the existence of such a
maximum accretion rate is not a generic property of advective flows.

\subsection{Slim disc solutions}

For $\alpha < \alpha_{\rm crit}$ there exists a separate branch of
solutions which represents for $\dot m \lta 1$ the standard 
Shakura-Sunyaev discs. Higher accretion rates are represented
by the so-called `slim discs' (Abramowicz et al. 1988). It is sometimes
said that slim discs represent solutions in which advection dominates
because the optical depths are so high that photons are trapped in
the flow. It is easy to see, however, that the optical depth of slim discs,
for a rather large range of parameters, is rather low. In fact, it
is the decrease of the optical depth with increasing accretion rate that
is at the origin of the slim disc branch of solutions. Slim discs are
only asymptotically advection dominated (see Fig 1. in Abramowicz et al.
1988).

\section{Global solutions}

The so called `self-similar' solutions found by Narayan \& Yi (1994)
played a very important role in the development of ADAF astrophysics
because they allowed Narayan and collaborators (see Narayan, Mahadevan
and Quataert 1998, for a recent review) to construct models which could
be quickly compared with observations. These comparisons showed that
ADAFs provide an excellent description of such systems as the Galactic
Center source Sgr A$^*$ (Narayan, Yi \& Mahadevan 1995) and quiescent
soft X-ray transients (SXTs) (Narayan, McClintock \& Yi 1995).  Of
course, these models were rather crude and not self-consistent (for
example, the `advection parameter' $f$, which must be constant for a
`self-similar' solution varied with radius) but after more refined and
consistent models had been calculated, they confirmed most of the
prediction of the `self-similar' ones. In fact, major revisions of the
early models concerned only the value of the transition radius between
the ADAF and the outer geometrically thin disc, whose presence is
required by observations of SXTs and the AGN NGC 4258 (see next
section).

The first 2T, global, optically thin ADAF solutions were obtained by
Matsumoto, Kato \& Fukue (1985). In Chen, Abramowicz \& Lasota (1997)
and Narayan, Kato \& Fukue (1997) one finds the first detailed study of
the properties of global ADAF solutions. The Narayan et al. (1997) work
deals with pure advection dominated solutions whereas Chen et al.
(1997) consider more general solutions which allow for bremsstrahlung
cooling.

\begin{figure}
\begin{center}
\epsfig{figure=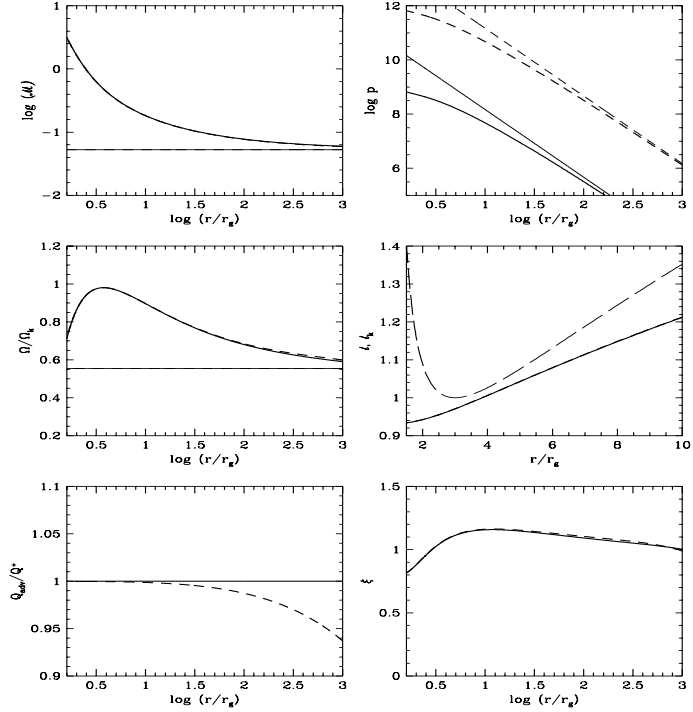,width=10cm,height=10cm 
}
\caption{Disk solutions for $\alpha=0.1$ and $\dot m =10^{-5}$ -
solid lines and $\alpha=0.1$ and $\dot m=10^{-2}$ - dashed lines.
The mass of the black hole is $10 M_{\odot}$.  
${\cal M}$ is the Mach number, $\Omega$ the angular frequency of
the flow and $\xi$ the advection parameter.
The heavy lines are the global solution and the thin lines are the
corresponding self-similar solution. Note that ${\cal M}$, $\Omega$,
and $\xi$ depend on $\dot m$ very weakly.}
\label{f1}
\end{center}
\end{figure}

\begin{figure}
\begin{center}
\epsfig{figure=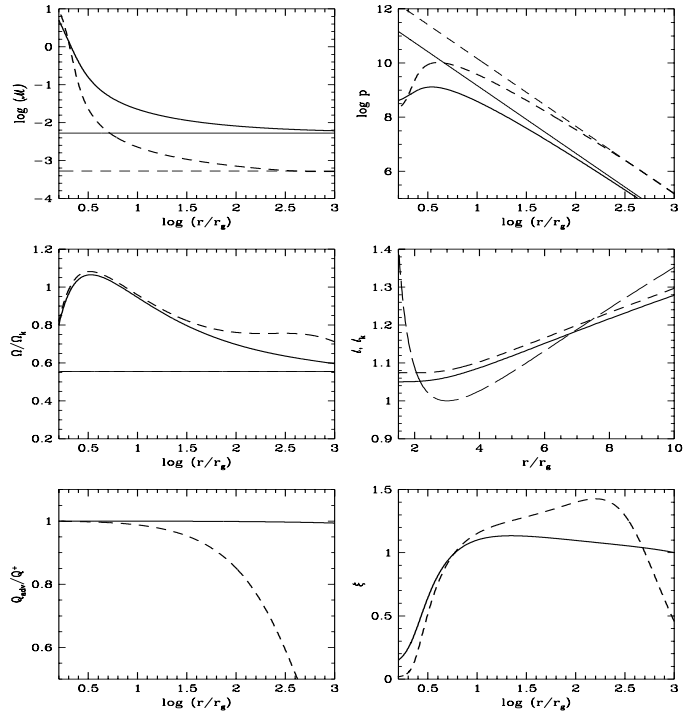,width=10cm,height=10cm 
}
\caption{Solutions for $\alpha=0.01$ and $\dot m =10^{-5}$ - solid
lines and $\alpha=0.001$ and $\dot m=10^{-5}$ - dashed lines.  
The black hole mass is the same as in Fig. (\ref{f1}).
Heavy
lines correspond to global solutions and thin lines correspond to
self-similar solutions.  Note the super-Keplerian angular momentum and
the maximum pressure near the transonic region.  Note also that for
$\alpha=0.001$ and $\dot m=10^{-5}$, the local cooling becomes
important for large radii.
}
\label{f2}
\end{center}
\end{figure}

Figs. (\ref{f1}) and (\ref{f2}) show the characteristic properties of
global ADAF solutions. They share some properties with slim disc
solutions, such as the sub-Keplerian character of the flow for `high'
values of $\alpha$ and a super-Keplerian part of the flow for low
values of this parameter. These features are related to the existence
of a maximum pressure in the flow.  Narayan et al. (1997a) argue that a
pressure maximum is necessary for the existence of `funnels' that
appear in `iron tori' (Rees et al. 1982).  Since no flow models for
iron tori exist it is difficult to tell the difference between these
structures and ADAFs.

Figs. (\ref{f1}) and (\ref{f2}) show also the influence of the maximum
accretion rate for ADAFs: for low $\alpha$'s and/or high $\dot m$'s a
flow which is advection dominated at small radii ceases to be an ADAF
at larger radii. 

\begin{figure}
\begin{center}
\epsfig{figure=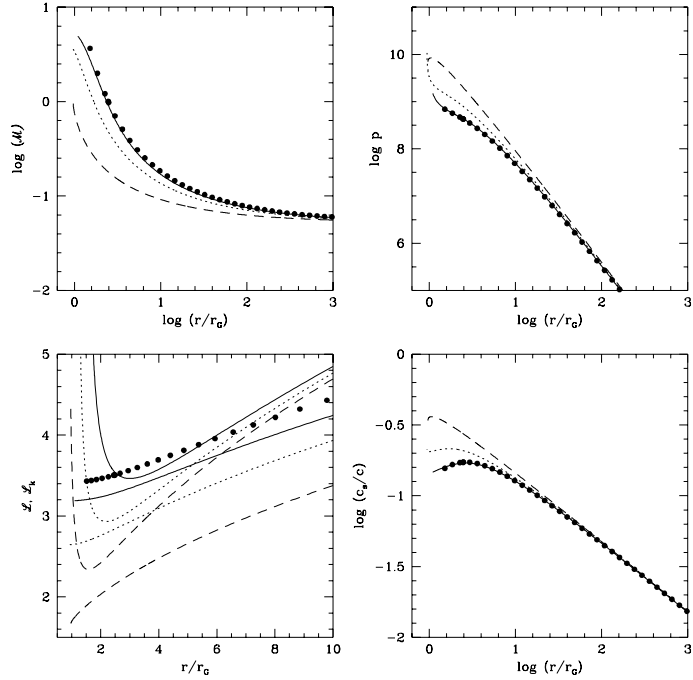,width=10cm,height=10cm 
}
\caption{The radial structure of the Mach number $({\cal M})$,
the pressure ($p$ in cgs unit), the angular momentum (${\cal L}$
in unit of $M$), and the sound speed ($c_s$ in unit of the speed
of light). The mass of the black hole is $10 M_{\odot}$,
$\alpha=0.1$ and $\dot M / \dot{M}_E = 10^{-5}$.  The solid,
dotted, and dashed lines represent the cases of $a/M=0$, 0.5,
and 0.99 respectively. The heavy dots represent solutions obtained
with the pseudo-Newtonian potential.  These solutions are excellent
approximation to the solutions representing  Schwarzschild black
hole flows (in the case $a=0$). The corresponding Keplerian angular
momenta of test particles around Kerr black holes are also shown for
comparison (the thin lines).}
\label{kerr}
\end{center}
\end{figure}

First studies of global ADAF solutions used the Paczy\'nski-Wiita
(1980) approximation of the black effective potential, but such an
approach is not satisfactory in the case of a rotating black hole.
After the first ADAF solutions in a Kerr spacetime were found by
Abramowicz et al. (1996), several other groups produced such ADAF
models (Peitz \& Appl 1997; Gammie \& Popham 1998; Popham \& Gammie
1998). Jaroszy\'nski \& Kurpiewski (1997) are the only authors who also
calculated ADAF spectra in the framework of General Relativity. Fig
(\ref{kerr}) show Abramowicz et al. (1996) solutions for three values
of the black hole angular momentum. One can see that these solutions
are very similar to the ``pseudo-relativistic" ones. In fact for a
non-rotating black hole the Paczy\'nski-Wiita (1980) ansatz is an
excellent approximation.

\section{Applications}

Despite several rather serious uncertainties about the physics of
ADAFs, their models are enjoying a growing field of application. The reason is simple.
In many astrophysical systems powered by accretion (see below), the X-ray
luminosity, which is supposed to probe accretion onto the central compact
object, is very weak compared with the expectations based on independent
estimates of the accretion rate and the assumption of high ($\sim 0.1$)
radiative efficiency. The only model which predicts such properties of
accreting systems is the ADAF model, of which a low radiative efficiency
is a fundamental feature.

Most models which reproduce observed spectra by ADAFs use
the approach pioneered by Ramesh Narayan and collaborators. It consists
in fixing `microscopic' parameters of the flow, such as the ratio of the
magnetic to gas pressure, the fraction of viscous dissipation that goes
directly to electrons, the $\alpha$-parameter, the thermodynamical
parameters, etc. Then the accretion rate  is determined by adjusting
to the observed X-ray flux. In a pure ADAF solution (no external disc)
this determines the whole spectrum so that no additional freedom is
allowed.  The `microscopic parameters' are considered to be universal so
this procedure is, in fact, an one-parameter `fit' (assuming of course
that the black hole mass is given by independent observations). In the
case of a two-component flow (ADAF + accretion disc) a second
`parameter' has to be fixed: the transition radius between the two
flows. In principle this should not be a free parameter, but should
be given by the physics of this transition. This is however a weak (the
weakest according to the present author) point of the ADAF `paradigm':
the mechanism of transition between the two flows is unknown. There
exist the `usual suspects', such as evaporation, but a consistent
physical model is yet to be found. A principle according to which
the flow will be an ADAF
when, at a given radius, an ADAF and an accretion disc solutions are
possible has been used (see e.g. Menou,
Narayan \& Lasota 1998) but the case of NGC 4258 seems to contradict
this principle (see Sect. 3.3).

\subsection{Sgr A$^*$}

Rees (1982) was the first to suggest that accretion onto the Galactic
Center black hole might be advection-dominated. Detailed models by
Narayan et al. (1995,1998) and Manmoto, Mineshige \& Kusunose (1997) provide
rather impressive fits to observations from radio to X-ray frequencies.
In this case the model is a pure ADAF; no outer accretion disc is present.

\subsection{Soft X-ray Transients and `Low state X-ray binaries'}

Soft X-ray Transients (SXTs) are a natural field for application of
ADAF models since  an accretion disc model is unable to fit the
observations of these systems in quiescence (Lasota 1996). In the first
ADAF model of quiescent SXTs proposed by Narayan, McClintock \& Yi
(1996) an outer stationary accretion disc was responsible for the
observed emission in optical and UV frequencies. It was shown however,
by Lasota, Narayan \& Yi (1996), that it is impossible to find an ADAF +
accretion disc configuration which would be consistent both with
observations and with the requirements of the disc instability model, which
is supposed to describe SXT outbursts. A new version of the model was
proposed by Narayan, Barret \& Yi (1997) in which the contribution of
the outer disc was negligible. This was achieved by increasing the
transition radius and increasing the magnetic to gas pressure ratio so
that the optical/UV emission is due to synchrotron radiation from the
ADAF.

The validity of this model has been independently confirmed by
Hameury et al. (1997). They reproduced the multi-wavelength properties
of the rise to outburst of GRO J1655-40, including the observed 6 day
delay between the rise in optical and X-rays, by using the disc
instability code of Hameury et al. (1998). As initial conditions they
used parameters given by an ADAF fit to the quiescent spectrum. They
also showed that it is very difficult (if not impossible) to reproduce
these observations with different initial conditions.

Narayan (1996) proposed to interpret various spectral and luminosity
states observed in X-ray binaries using ADAF + accretion disc models
with varying accretion rate and transition radius. This idea has been
applied to the black hole SXT X-ray Nova Muscae 1991 (Esin et al. 1997)
and to Cyg X-1 (Esin et al. 1998; the first to apply an ADAF model to
this system was Ichimaru 1984). The existence of a two-component
flow with varying accretion rate and transition radius in the 
XNova Muscae 1991 was confirmed by observations of the X-ray reflected
component in the Ginga spectra of this object, but these observations
suggest (\.Zycki, Done \& Smith 1998) that the Esin et al. (1997) model
requires some modifications.

\subsection{NGC 4258}

This LINER is a very important testing ground for the ADAF models. The
black hole mass in this system ($3.6 \times 10^7 M_{\odot}$) is very
well determined due to the presence of narrow, water maser lines
(Miyoshi et al. 1997).  The observed X-ray luminosity is $\sim 10^{-5}
L_{\rm Eddington}$ and the bolometric luminosity is no more than an
order of magnitude larger. Lasota et al. (1996b) proposed, therefore, that
the inner accretion in this system proceeds through an ADAF. New
observations in infrared (Chary \& Becklin 1997) and a new upper limit on
the radio flux (Herrnstein et al. 1998) constrain the transition radius
to be $r_{\rm tr} \sim 30$ (Gammie, Blanford \& Narayan 1998). In this
model the accretion rate would be $\dot m= 9 \times 10^{-3}$, much
higher than the value of $\dot m \sim 10^{-5}$ proposed by Neufeld \&
Maloney (1995) in their model of the masing disc, but in agreement with
values obtained by Maoz \& McKee (1997) and Kumar (1997) for the same
disc. It seems that the value proposed by  Neufeld \& Maloney (1995) is
excluded by the IR and X-ray data. In NGC 4258, an ADAF solution is possible
for $r>r_{\rm tr}$ so that the validity of the principle according to
which an ADAF is preferred over an accretion disc whenever the two
solutions are possible seems  questionable.

\subsection{LINERs and weak AGNs}

Lasota et al. (1996b) suggested that also other LINERS and `weak' AGNs could
contain ADAFs in their inner accretion regions. A recent analysis
of the variability of such systems (Ptak et al. 1998) brings new arguments
in favour of this hypothesis.

\subsection{Nuclei of giant elliptical galaxies}

Fabian \& Rees (1995) proposed that the well known problems with understanding
the properties of nuclei of giant elliptical galaxies, which emit much fewer
X-rays than one would expect from independent estimates of the accretion rate,
could be solved if the accretion flow formed an ADAF. Reynolds et al. (1996)
and di Matteo \& Fabian (1997) applied  ADAF models to the nuclei of M87 and
M60 respectively. The latter authors point out that observations
of the predicted spectral turnover at $\sim 3 \times 10^{11}$ Hz puts
constrains on the values of the `microscopic' parameter of the ADAF.

\subsection{BL Lac's}

In BL Lac's the emission is dominated by the jet but several properties of
these systems points out to the possible presence of an ADAF (Madejski 1996;
Celotti, Fabian \& Rees 1997).

\subsection{Remnant and MACHO black holes in the ISM}

Fujita et al. (1998) noticed that if black holes accreting from the interstellar
medium formed an ADAF, the characteristic ADAF spectral form would help 
in their detection.

\section{Conclusion}

ADAF models find successful applications in many domains of accretion
astrophysics.  They are the only models which describe both the
dynamics of the accretion flow and its emission properties. In any
case, all models which require the presence of optically thin, very hot
plasmas (such as `coronal' models) must take into account advective
heat transport in order to be self-consistent (see Section 1.2).  ADAFs
are the only solutions that satisfy this requirement.

Of course, as mentioned above, there are still serious problems to be
solved before one can conclude that ADAFs, in their present form, are
the models of accretion flows around black holes at low accretion
rates. The main problem to be solved is that of the transition between
an accretion disc and an ADAF.  Furthermore, in order to reproduce
data, ADAF should form two-temperature plasmas with ions at virial
temperature and much cooler electrons. This requires weak coupling
between ions and electrons and implies that viscosity heats mainly the
ions. Some recent studies suggest that it is difficult to achieve this
together with equipartition between magnetic and gas pressures
(Quataert \& Gruzinov 1998). Such calculations however, concern
extremely complicated physical processes so that it is not clear that
results of numerical simulations apply to real systems.  One should
therefore apply a more pragmatic attitude, that the apparently
successful applications of ADAFs to many astrophysical systems suggests
that no mechanisms coupling ion and electron exist in real
astrophysical configurations (Fabian \& Rees 1995).

The validity of ADAF models will be decided by observations.

\section*{Acknowledgements}

I am grateful to Charles Gammie for very interesting discussions.  I
thank the Isaac Newton Institute in Cambridge for hospitality in
April 1998 when most of this article was written.

\section*{References}

\noindent
Abramowicz, M.A., Chen, X.M., Kato, S., Lasota, J.-P. \& Regev, O. 1995,
ApJ, 438, L37\\
\noindent
Abramowicz, M.A., Chen, X.M., Granath, M. \& Lasota, J.-P. 1996,
 ApJ, 471, 762\\
\noindent
Bj\"ornsson, G., Abramowicz, M.A., Chen, X.M. \& Lasota, J.-P. 1996, ApJ,
467, 99\\
\noindent
Celotti, A., Fabian, A.C. \& Rees, M.J. 1997, MNRAS, 293, 239\\
\noindent
Chary, R. \& Becklin, E.E. 1997, ApJ, 485, L75\\
\noindent
Chen, X.M., Abramowicz, M.A. \& Lasota, J.-P. 1997, ApJ, 476, 61\\
\noindent
Chen, X.M., Abramowicz, M.A., Lasota, J.-P., Narayan, R. \& Yi, I. 1995,
ApJ, 443, L61\\
\noindent
di Matteo, T. \& Fabian, A.C. 1997, MNRAS, 286, 393\\
\noindent
Esin, A.A. 1997, ApJ, 482, 400\\
\noindent
Esin, A.A., McClintock, J.E., \&Narayan, R. 1997, ApJ, 489, 865\\
\noindent
Esin, A.A., Narayan, R., Ostriker, E. \& Yi, I. 1996, ApJ, 465, 312\\
\noindent
Esin, A.A., Narayan, R., Cui, W., Grove, J.E. \& Zhang, S-N. 1998, 
ApJ, in press\\
\noindent
Fabian, A.C. \& Rees, M.J. 1995, MNRAS, 277, L5\\
\noindent
Fujita, Y., Inoue, S., Nakamura, T., Manmoto, T. \& Nakamura, K.E. 1998, ApJ,
495, 85\\
\noindent
Gammie, C.F. \& Popham, R. 1998, ApJ, in press\\
\noindent
Gammie, C.F., Blanford, R.D. \& Narayan, R. 1998, ApJ, in preparation\\
\noindent
Hameury, J.-M., Lasota, J.-P., McClintock, J.E. \& Narayan, R. 1997, 
ApJ, 489, 234\\
\noindent
Hameury, J.-M., Menou, K., Dubus, G. \& Lasota, J.P. 1998, MNRAS, in press\\
\noindent
Herrnstein, J. et al. 1998, ApJ, 497, L73\\
\noindent
Ichimaru 1984\\
\noindent
Kumar, P. 1998, ApJ, submitted, astro-ph/9706063\\
\noindent
Jaroszy\'nski, M. \& Kurpiewski, A. 1997, A\&A, 326, 419\\
\noindent
Lasota, J.P. 1996, in {\it Compact Stars in Binaries}, IAU Symp. 165,
eds J. van Paradijs, E.P.J. van den Heuvel \& E. Kuulkers, KLUWER Academic Pub. 
Dordrecht, The Netherlands. p. 43\\
\noindent
Lasota, J.-P., Narayan, R. \& Yi, I. 1996a, A\&A, 314, 813\\
\noindent
Lasota, J.-P., Abramowicz, M.A., Chen, X.M., Krolik, J. Narayan, R. \& Yi, I. 
1996b, ApJ,  462, 142\\
\noindent
L\o v\aa s, T. 1998, Master's Thesis, Copenhagen University Observatory\\
\noindent
Madejski, G. 1996, private communication\\
\noindent
Manmoto, T., Mineshige, S. \& Kusunose, M. 1997, 489, 791\\
\noindent
Maoz \& McKee
\noindent
Matsumoto, R., Kato, S. \& Fukue, J. 1985, in {\it Theoretical Aspects of
Structure and Evolution of Galaxies}, eds S. Aoki, M. Iye \& Y. Yoshii
Tokyo Observatory, p. 102\\
\noindent
Miyoshi, M. et al. 1995, Nature, 373, 127\\
\noindent
Menou, K., Narayan, R. \& Lasota, J.-P. 1998, ApJ, submitted\\
\noindent
Nakamura, K.E., Kususnose, M., Matsumoto, R. \& Kato, S. 1997,
PASJ, 49, 503\\
\noindent
Narayan, R. 1996, ApJ, 462, L13\\
\noindent
Narayan, R. \& Yi, I. 1994, ApJ, 428, L13\\
\noindent
Narayan, R. \& Yi, I. 1995, ApJ, 452, 710\\
\noindent
Narayan, R., Yi, I. \& Mahadevan, R. 1995, Nature, 374, 623\\
\noindent
Narayan, R., McClintock, J.E. \& Yi, I. 1996, ApJ, 457, 821\\
\noindent
Narayan, R., Kato, S. \& Honma, F. 1997a, ApJ, 476, 49\\
\noindent
Narayan, R., Barret, D. \& Yi, I. 1997b, ApJ, 482, 448\\
\noindent
Narayan, R., Mahadevan, R., Grindlay, J.E., Popham, R.G. \& Gammie, C.F. 1998,
492, 554\\
\noindent
Neufeld, D.A. \& Maloney, P.R. 1995, ApJ. 447, L17\\
\noindent
Paczy\'nski, B. \& Wiita, P.J. 1980, A\&A, 88, 23\\
\noindent
Peitz, J. \& Appl, S. 1997, MNRAS, 286, 681
\noindent
Popham, R. \& Gammie, C.F. 1998, ApJ, submitted\\
\noindent
Ptak, A., Yaqoob, T., Mushotzky, P., Serlemitsos, P. \& Griffiths,  R. 1998,
ApJL, in press, astro-ph/9804327\\
\noindent
Quataert, E. \& Gruzinov, A. 1998, ApJ, submitted, astro-ph/9803112\\
\noindent
Rees, M. J. 1982, in {\it Galactic Center}, eds G.R. Riegler \& R.D. Blanford,
AIP, New York, USA, p. 166\\
\noindent
Rees, M. J., Begelman, M.C., Blanford, R.D. \& Phinney, E.S. 1982, Nature,
295, 17\\
\noindent
Reynolds, C.S., Di Matteo, T., Fabian, A.C.,  Hwang, U. \& Canizares, C.R.
1996, MNRAS, 283, L111\\
\noindent
\.Zycki, P.T., Done, C. \& Smith, D.A. 1998, ApJ, 496, L25\\
\noindent 

\newpage

\end{document}